\documentclass[conference]{IEEEtran}

\IEEEoverridecommandlockouts
\def\BibTeX{{\rm B\kern-.05em{\sc i\kern-.025em b}\kern-.08em
    T\kern-.1667em\lower.7ex\hbox{E}\kern-.125emX}}

\AtBeginDocument{%
  \providecommand\BibTeX{{%
    Bib\TeX}}}

\usepackage{pervasives}
\usepackage{balance}

\begin{document}

\title{A Framework for Fine-Grained Synchronization of Dependent GPU Kernels}

\author{\IEEEauthorblockN{Abhinav Jangda}
\IEEEauthorblockA{Microsoft Research\\
                  United States}
\and
\IEEEauthorblockN{Saeed Maleki}
\IEEEauthorblockA{Microsoft Research\\
                  United States}
\and
\IEEEauthorblockN{Maryam Mehri Dehnavi}
\IEEEauthorblockA{University of Toronto\\
                  Canada}
\and
\IEEEauthorblockN{Madan Musuvathi}
\IEEEauthorblockA{Microsoft Research\\
                  United States}
\and
\IEEEauthorblockN{Olli Saarikivi}
\IEEEauthorblockA{Microsoft Research\\
                  United States}
}

\maketitle

\begin{abstract}
Machine Learning (ML) models execute several parallel computations including Generalized Matrix Multiplication, Convolution, Dropout, etc.
These computations are commonly executed on Graphics Processing Units (GPUs), by dividing the computation into independent processing blocks, known as \emph{tiles}.
Since the number of tiles are usually higher than the execution units of a GPU, tiles are executed on all execution units in one or more \emph{waves}.
However, the number of tiles is not always a multiple of the number of execution units. 
Thus, tiles executed in the final wave can under-utilize the GPU.

To address this issue, we present \libname{}, a framework for synchronizing dependent kernels using a user-defined fine-grained synchronization policy to improve the GPU utilization.
\libname{} synchronizes tiles instead of kernels, which allows executing \emph{independent} tiles of \emph{dependent} kernels concurrently.
We also present a compiler to generate diverse fine-grained synchronization policies based on dependencies between kernels.
Our experiments found that synchronizing CUDA kernels using \libname{} reduces the inference times of four popular ML models: MegatronLM GPT-3 by up to 15\%, LLaMA by up to 14\%, ResNet-38 by up to 22\%, and VGG-19 by up to 16\% over several batch sizes.
\end{abstract}

\begin{IEEEkeywords}
CUDA, GPU, Generalized Matrix Multiplication, Convolution, Fine-Grained Synchronization, Machine Learning
\end{IEEEkeywords}

\section{Introduction}
The trend of larger Machine Learning (ML) models has delivered remarkable results in multiple domains.
These results have exploded the demand of ML models in innumerable applications.
To serve this demand, the infrastructure for running inference on these large models has also scaled up exponentially.
Hence, optimizing for even the last percentage in the inference can lead to huge savings in cost and energy of serving these models.

ML models are typically served using multiple GPUs because these models consist of embarrassingly parallel operations, such as Generalized Matrix Multiplication (GeMM), 2-D Convolution (Conv2D) etc.
The traditional approach to execute a computation on a GPU breaks down the computation into multiple independent blocks, known as \emph{tiles}.
Each tile is computed by a fixed size block of threads, known as a \emph{thread block}, which runs on an execution unit of the GPU known as a \emph{Streaming Multiprocessor} (SM).
Often the number of thread blocks are higher than the number of SMs.
Therefore, all thread blocks are executed in one or more \emph{waves}, with initial full waves executing thread blocks that are a multiple of the number of SMs and the final partial wave executing less than the number of SMs thread blocks.
When executing a pair of dependent operations, the traditional approach executes these operations on the same \emph{stream}.
Executing two or more operations on a stream, ensures that no thread block of a later operation can execute before the thread blocks of all former operations are finished.
We call this traditional heavy-weight synchronization approach as \emph{stream synchronization}.

\begin{figure*}[ht]
  \centering
  \begin{subfigure}[b]{0.31\textwidth}
    \includegraphics[scale=0.6]{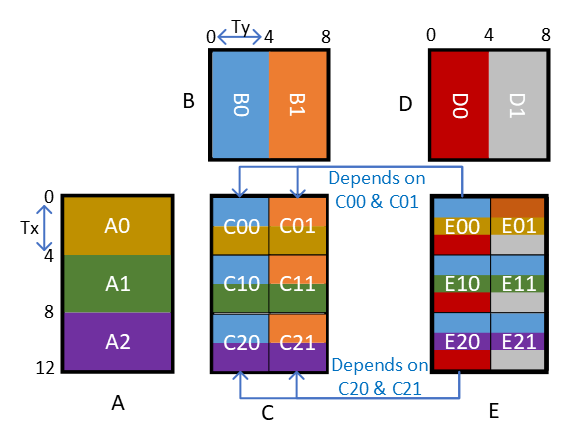}
    \caption{\textit{Tiled GeMM kernels}: A tile $C_{i,j}$ is computed by multiplying sub-matrices $A_{i}$ and $B_{j}$.
    Similarly, a tile $E_{i,j}$ is computed by multiplying $C_{i}$ with $D_{j}$.
    Since each tile is computed by one thread block, the tile size of $4\times 4$ gives the grid size of $\{\frac{12}{4}, \frac{8}{4}\} = \{3, 2\}$ for both kernels.
    Both kernels have the occupancy of 1 thread block per SM. 
      \label{fig:motivation:tiled-matmul}} 
  \end{subfigure}
  \hfill{}
  \begin{subfigure}[b]{0.31\textwidth}
  \includegraphics{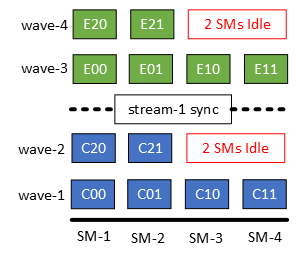}
  \caption{
    \textit{Stream Synchronization} synchronizes all thread blocks of both kernels.
  Thread blocks of both producer ($C_{i,j}$) and consumer ($E_{i,j}$) are executed in two waves.
  The first wave executes four thread blocks and the second wave executes remaining two.
  In the second wave of both kernels, SM-3 and SM-4 are not utilized.
  \label{fig:motivation:current-scenario}}
  \end{subfigure}
  \hfill{}
  \begin{subfigure}[b]{0.32\textwidth}
  \includegraphics[scale=0.97]{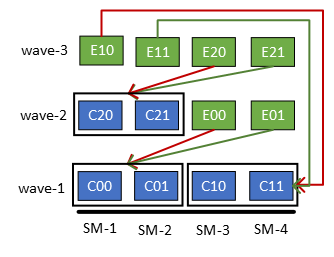}
  \caption{
    \textit{Fine-grained Synchronization} synchronizes only dependent thread blocks (shown as arrows) of both kernels and executes in only three waves.
  Thread blocks of the consumer-kernel waits using a semaphore until its producer-kernel's thread block has computed the dependent tile. 
  Since in every wave all SMs are utilized, we achieve full utilization.
  \label{fig:motivation:fine-grained synchronization}}
  \end{subfigure}
  \caption{Thread block execution with existing stream synchronization and fine-grained synchronization on 4 SMs for two dependent GeMM kernels: $C_{12\times 8} = A_{12\times 8} \times B_{8\times 8}$ and $E_{12\times 8} = C_{12\times 8} \times B_{8\times 8}$.\label{fig:motivation}}
\end{figure*}

However, this heavy-weight synchronization can lead to the under-utilization of GPU resources in the final wave when thread blocks are not a multiple of SMs.
For example, Figure~\ref{fig:motivation:tiled-matmul} shows that executing 6 tiles of two dependent GeMMs on four SMs require $\lceil\frac{6}{4}\rceil = 2$ waves for each GeMM.
With stream synchronization, no thread block of the second GeMM can execute before all thread blocks of the first GeMM are finished. 
Thus, as Figure~\ref{fig:motivation:current-scenario} shows, the second partial wave of each GeMM utilizes only two out of four SMs.
This under-utilization is prevalent in widely used ML models.
Table~\ref{tab:intro} shows that during the inference of MegatronLM GPT-3~\cite{megatronlm}, the two dependent GeMMs achieves 60--80\% of utilization on an NVIDIA Tesla V100 GPU because the number of thread blocks are not a multiple of the number of SMs.

\begin{table}[t]
  \vspace{1em}
  \caption{\sc Number of thread blocks (TBs), thread blocks per wave, waves, and GPU utilization of two dependent GeMMs in MegatronLM GPT-3~\cite{megatronlm} on several batch sizes when executing on an NVIDIA Tesla V100 containing 80 SMs.
  \label{tab:intro}}
  \centering
  \begin{tabular}{|r|r|r|r|r|r|r|}
    \hline
    \textbf{Batch} & \textbf{GeMM} & \textbf{TBs} & \textbf{\thead{TBs per\\Wave}} &\textbf{Waves} &\textbf{\thead{Utili-\\zation}}\\
    \hline
    \multirow{2}{*}{256}& Producer & [1, 48, 4] & 2$\times$80 & 1.2  & 60\%\\ 
                        & Consumer & [1, 96, 2] & 2$\times$80 & 1.2  & 60\%\\
    \hline
    \multirow{2}{*}{512}& Producer & [2, 24, 2] & 1$\times$80 & 1.2  & 60\%\\
                        & Consumer & [2, 48, 1] & 1$\times$80 & 1.2  & 60\%\\
    \hline
    \multirow{2}{*}{1024}& Producer & [4, 24, 2] & 1$\times$80 & 2.4  & 80\%\\
                         & Consumer & [4, 48, 1] & 1$\times$80 & 2.4  & 80\%\\
    \hline
  \end{tabular}
\end{table}

The state-of-the-art technique for executing GeMM computations on GPUs, Stream-K~\cite{streamk}, can improve the utilization of the final wave of a workload by partitioning tiles of the final wave among multiple thread blocks.
However, Stream-K suffers from three issues. 
First, partitioning a tile among multiple thread blocks requires each thread block to update the tile elements, leading to extra global memory accesses.
Second, Stream-K requires different kernel invocations for initial full waves and for the final partial wave.
Third, it is not straightforward to extend Stream-K's approach to other tile based computations including Dropout and Softmax.

In this paper, we present several \emph{fine-grained synchronization} techniques for synchronizing tiles of dependent computations enabling the execution of \emph{independent} tiles of both computations concurrently in the final wave.
Figure~\ref{fig:motivation:fine-grained synchronization} shows how one of our techniques, tile synchronization, obtains full utilization in our example.
We invoke both kernels on separate streams 
and synchronize only the dependent tiles, thus thread blocks, using a semaphore stored in the GPU memory.
Therefore, thread blocks of both kernels are executed in only three waves, leading to full utilization of the GPU.
However, as we show in the paper, the granularity of synchronization that provides the best performance depends on computations, data sizes, and GPU architecture.
To this end, we propose, \libname{}, a framework to efficiently synchronize dependent computations based on user-defined synchronization policies.
\libname{} contains mechanisms to:
(i) ensure that all thread blocks of the producer are executed before the consumer (Section~\ref{sec:stage-order}),
(ii) allow processing of producer and consumer tiles in an order that minimizes the wait time of synchronization by consumer tiles (Section~\ref{sec:tile-order}), 
(ii) maintain the dependence between tiles of producer and consumer computations using semaphores and memory fences (Section~\ref{sec:dependent-tiles}).
Furthermore, we propose a DSL to describe dependencies between GPU kernels and a compiler \compiler{} to generate synchronization policies from the DSL specification for \libname{} (Section~\ref{sec:cusyncgen}).
We described dependencies between computations of several ML models in the DSL and generated synchronization policies for diverse GPU computations, such as GeMM, 2-D Convolutions, and Dropout, using \compiler{}.
Synchronizing GPU computations using \libname{} reduces the inference time of several state-of-the-art open source ML models on 8x NVIDIA Tesla V100 GPUs: MegatronLM GPT-3 145 Billion parameter model~\cite{megatronlm} by 6--15\%, LLaMA 65.2 Billion parameter model~\cite{touvron2023llama} by 9--14\%, ResNet-38~\cite{resnet} by 5--22\%, and VGG-19~\cite{vgg} by 6--16\% (Section~\ref{sec:eval}).

\section{Background}
This section provides a background on NVIDIA GPUs and ML models.

\subsection{NVIDIA Graphics Processing Units and CUDA}
A parallel computation executing on NVIDIA GPUs is called a CUDA kernel.
A CUDA kernel executes multiple concurrent \emph{threads} organized in a 3-dimensional \emph{grid}, and these threads are grouped into equally sized \emph{thread blocks}.
The \texttt{dim3} struct in CUDA represents a 3-D grid size and identifier for both threads and thread blocks in \texttt{x}, \texttt{y}, and \texttt{z} dimensions.  
An NVIDIA GPU contains multiple Streaming Multiprocessors (SMs), each of which executes one or more thread blocks.
This number of thread blocks per SM, known as \emph{occupancy}, depends on the register and shared memory usage, thread block size, and number of thread blocks of the CUDA kernel. 

\spara{Thread block Wave Execution}
Thread blocks are executed on all SMs in $\lceil \frac{Number\_of\_TBs\_in\_Grid}{occupancy \times Number\_of\_SMs} \rceil$ \emph{waves}, where the \emph{initial full waves} execute $occupancy \times Number\_of\_SMs$ thread blocks and the \emph{final partial wave} execute the remaining thread blocks.
NVIDIA has not documented the mechanism for scheduling thread blocks to SMs followed by CUDA and GPUs.


\spara{Stream Synchronization}
A CUDA \emph{stream} is a sequence of CUDA operations that execute in the order they were issued.
When two dependent kernels are invoked on the same stream,
the consumer-kernel is not started before all thread blocks of the producer-kernel have finished their execution. 
We call this synchronization \emph{stream synchronization}.
We can invoke independent CUDA kernels on different streams to execute kernels concurrently.
A stream has an associated priority value, such that operations on a higher priority stream are issued before a lower priority stream.

\subsection{Computations in Large ML Models}
Contemporary ML models contain embarrassingly parallel computations, such as, Generalized Matrix Multiplication (GeMM), 2-D Convolution (Conv2D), Dropout, and Softmax.
We consider four widely used machine learning models: MegatronLM GPT-3 145B ~\cite{megatronlm}, LLaMA 65.2B~\cite{touvron2023llama}, ResNet-38~\cite{resnet}, and VGG-19~\cite{vgg}.
Below we briefly explain computations involved in these models.
 
\begin{figure}[t]
\begin{subfigure}[b]{\columnwidth}  
\begin{lstlisting}[language=CUDA, xleftmargin=1.0em]
//X: [B,S,H]; W|\color{commentgreen}{$_1$}|: [H,4H/8]; W|\color{commentgreen}{$_2$}|: [4H/8,H]
//1st GeMM fused with GeLU XW|\color{commentgreen}{$_1$}|: [B,S,4*H/8]
XW|$_1$| = GeLU(X |$\times$| W|$_1$|)|\label{line:mlp:first-matmul}|
//2nd GeMM XW|\color{commentgreen}{$_2$}|: [B,S,H]
XW|$_{12}$| = XW|$_1$| |$\times$| W|$_2$|
\end{lstlisting}
\caption{Multi-Layer Perceptron (MLP) contains two weight matrices: \texttt{W$_1$} of shape $\left[\texttt{H}, \frac{4\texttt{H}}{8}\right]$ and \texttt{W$_2$} of shape $\left[\frac{4\texttt{H}}{8}, \texttt{H}\right]$.
\label{fig:gpt-3:mlp}}
\end{subfigure}
\begin{subfigure}[b]{\columnwidth}
\begin{lstlisting}[language=CUDA, xleftmargin=1.0em]
//X: [B,S,H]; QKV: [H,3H/8]; W|\color{commentgreen}{$_2$}|: [H/8,H]
//1st GeMM XQKV: [B,S,3H/8]
XQKV = X |$\times$| QKV
//XQ: [B,S,H/8]; XK: [B,S,H/8]; XV: [B,S,H/8]
XQ = XQKV[:,:,0:H/8]    //1st matrix slice
XV = XQKV[:,:,H/8:2*H/8]//2nd matrix slice
XK = XQKV[:,:,2*H/8:]   //3rd matrix slice
//Cached Attention Mechanism
//CachedK: [H/8,S',B  ]
//CachedV: [B  ,S',H/8]
P = XQ |$\times$| Concat(CachedK, XK.T) |\label{line:attn:attention-start}|
R = Softmax(Dropout(P))
T = R |$\times$| Concat(CachedV, XV) |\label{line:attn:attention-end}|
CachedV[:S'+S:] = XV
CachedK[:S'+S:] = XK.T
//2nd GeMM XW|\color{commentgreen}{$_2$}|: [B,S,H]
XW|$_{12}$| = R |$\times$| W|$_2$|
\end{lstlisting}
\caption{Attention contains two weight matrices: \texttt{QKV} of shape $\left[\frac{3\texttt{H}}{8}, \texttt{H}\right]$, and \texttt{W$_2$} of shape $\left[\frac{\texttt{H}}{8}, \texttt{H}\right]$.
Attention caches generated keys and values for each token to avoid recomputation of all previous tokens during inference.  
\label{fig:gpt-3:attention}}
\end{subfigure}
\caption{Architecture of Multi-Layer Perceptron (MLP) and Attention of GPT-3, where H is 12288.
Model parallelism on 8 GPUs divides weight matrices of both layers among 8 GPUs.
Both takes an input matrix \texttt{X} of shape $\left[\texttt{B}, \texttt{S}, \texttt{H}\right]$ and obtain the result \texttt{XW$_{12}$} of the same shape.
\texttt{B} is the number of batched requests, \texttt{S} is the sequence length, \texttt{H} is the hidden dimension, and \texttt{S'} is the sum of processed and generated tokens.
\label{fig:gpt-3}}
\end{figure}

\begin{figure}[t] 
\begin{lstlisting}[language=CUDA, xleftmargin=1.0em]
//X: [B, S, H]; W|\color{commentgreen}{$_1$}|: [H, H/3];
//V: [H, H/3]; W|\color{commentgreen}{$_2$}|: [H/3, H]
//1st GeMM XW|\color{commentgreen}{$_1$}|: [B, S, H/3] 
XW|$_1$| = GeLU(X |$\times$| W|$_1$|)|\label{line:mlp:first-matmul}|
//2nd GeMM XV: [B, S, H/3]
XV = X |$\times$| V |\label{line:mlp-llama:first-matmul}|
//SwiGLU fused with 3rd GeMM  XW|\color{commentgreen}{$_2$}|: [B, S, H]
SwiGLU = Swish(XW|$_1$|) $\cdot$ XV
XW|$_{12}$| = SwiGLU |$\times$| W|$_2$|
\end{lstlisting}
\caption{The LLaMA MLP contains three weight matrices. With model parallelism on 8 GPUs, these matrices are: \texttt{W$_1$} of shape $\left[\texttt{H}, \frac{\texttt{H}}{3}\right]$, \texttt{V} of shape $\left[\texttt{H}, \frac{\texttt{H}}{3}\right]$, and \texttt{W$_2$} of shape $\left[\frac{\texttt{H}}{3}, \texttt{H}\right]$.
\label{fig:llama:mlp}}
\end{figure}
\subsubsection{Transformers Models} 
A \emph{transformer} is a deep learning architecture for Natural Language Tasks and is the basis of two widely used models: MegatronLM GPT-3~\cite{megatronlm} and LLaMA~\cite{touvron2023llama}.
An inference request to a transformer model consists of a prompt and is served in two phases: (i) \emph{prompt processing}, where the prompt is processed, and (ii) \emph{token generation}, where a series of tokens that represents the output response text is generated incrementally.
The model can batch \texttt{B} requests into a single inference task.
The sequence length \texttt{S} denotes the number of tokens of each request being processed in the prompt processing phase or the number of tokens of each request being generated in the token generation phase. 
Therefore, during prompt processing $\texttt{B} \geq 1, \texttt{S} > 1$ and during token generation $\texttt{B} \geq 1, \texttt{S} = 1$.

A transformer consists of multiple Multi-Layer Perceptron (MLP) and Attention blocks.
The design of MLP and Attention can be different for each model.

\spara{GPT-3}: In GPT-3, both MLP and Attention takes an input matrix, perform operations with its two weight matrices, and outputs a matrix.
With model parallelism these weight matrices are divided among all GPUs~\cite{megatronlm}.
Figure~\ref{fig:gpt-3} shows computations of GPT-3 with model parallelism of 8 GPUs.
Both MLP and Attention first applies a linear transformation on the input, i.e., perform GeMM of the input and the weight matrix.
Then, they perform operations, such as, GeLU and the Attention mechanism.
Finally, the output of this operation is applied to second linear transformation.
Existing MLP implementations fuse the GeLU activation with the first GeMM (line~\ref{line:mlp:first-matmul} in Figure~\ref{fig:gpt-3:mlp}).
State-of-the-art Attention implementations~\cite{flash-attention} caches the already processed and generated tokens in a KV Cache, such that, after prompt processing number of cached tokens, i.e. \texttt{S'}, is set to \texttt{S} and when generating tokens \texttt{S'} increases incrementally.
These implementations also fuses the attention mechanism in a single CUDA kernel (line~\ref{line:attn:attention-start}--line~\ref{line:attn:attention-end}) in Figure~\ref{fig:gpt-3:attention}.

\spara{LLaMA}: LLaMA uses the hidden dimension of size 8192.
LLaMA's MLP contains three GeMMs and SwiGLU~\cite{shazeer2020glu} activation as shown in Figure~\ref{fig:llama:mlp}.
State-of-the-art implementations combines first two GeMMs into a single GeMM and
fuses the SwiGLU activation with the third GeMM.
Moreover, LLaMA uses the same Attention architecture as GPT-3.

\subsubsection{Computer Vision Models} 
ResNet-38~\cite{resnet} and VGG-19~\cite{vgg} are two state-of-the-art computer vision models, where each layer performs several Conv2D operations.
Table~\ref{tab:resnet-sizes} shows the details of each convolution layer.


\begin{table}[t]
  \caption{\sc Input/output image size (\texttt{P}, \texttt{Q}, \texttt{C}), kernel size (\texttt{R}, \texttt{S}), channels (\texttt{K}) for each Conv2D, number of Conv2Ds per layer, and number of layers in ResNet-38 and VGG-19.\label{tab:resnet-sizes}}
  \centering
  \footnotesize
  \begin{tabular}{|r|r|r|r|r|r|r|r|r|r|}
    \hline
    \textbf{[\texttt{P}, \texttt{Q}, \texttt{C}]}& \textbf{[\texttt{R}, \texttt{S}]} & \textbf{\texttt{K}} & \multicolumn{2}{c|}{\textbf{Convs/Layer}} & \multicolumn{2}{c|}{\textbf{Layers}}\\
    \cline{4-5} \cline{6-7}
    & & &  \textbf{ResNet} & \textbf{VGG} & \textbf{ResNet} & \textbf{VGG}\\
    \hline
    $\threeDShape{56}{56}{64}$ & $\twoDShape{3}{3}$ & 64  & 2 & 2 & 3 & 1 \\
    $\threeDShape{28}{28}{128}$ & $\twoDShape{3}{3}$ & 128 & 2 & 2 & 4 & 1 \\
    $\threeDShape{14}{14}{256}$ & $\twoDShape{3}{3}$ & 256 & 2 & 4 & 6 & 1 \\
    $\threeDShape{7}{7}{512}$   & $\twoDShape{3}{3}$ & 512 & 2 & 4 & 3 & 1\\
    \hline
  \end{tabular}
\end{table}

\section{Fine-Grained Synchronization of Kernels}  
Our fine-grained synchronization of dependent CUDA kernels consists of four novel  mechanisms.
These mechanisms (i) ensure simultaneous allocation of dependent kernels (Section~\ref{sec:invoke-kernel}), (ii) execute thread blocks of producer kernels before consumer kernels (Section~\ref{sec:stage-order}), (iii) control the order of tile processing in each kernel to minimize the wait time of synchronization (Section~\ref{sec:tile-order}), and (iv) performs fine-grained synchronization of only dependent tiles of producer and consumer kernels (Section~\ref{sec:dependent-tiles}).
We have implemented these mechanisms in a header-only standalone CUDA library, \libname{}.

Figure~\ref{fig:ex:matmuls} explains these mechanisms using an example of synchronizing the two dependent GeMM kernels of MLP using \libname{}. 
The {\tt gemm} function is the standard GeMM GPU kernel (we use NVIDIA CUTLASS~\cite{cutlass} for our experiments) with additional code (shown as \underline{underlined}) to call into \libname{}. 
\libname{} associates each kernel with a {\tt CuStage} object that provide synchronization facilities among kernels.
The {\tt MLP} function creates these stage objects, declares dependencies between them, and invokes kernels.

\begin{figure*}[ht]
  \begin{subfigure}[b]{0.5\textwidth}
  \begin{lstlisting}[language=CUDA, xleftmargin=3.0ex]
//CUDA Kernel to compute C = A * B
global void gemm(f16* A, f16* B, f16* C, |\label{line:ex:kernel-start}| 
                    int K, |\underline{CuStage stage}|) {
  |\underline{stage.start();}| |\label{line:ex:init-prod-cons}||\underline{row, col = stage.tile();}| |\label{line:ex:get-tile}|
  for (tk = 0; tk < K; tk += TileK) {
    |\underline{stage.wait({A, row, tk});}| |\label{line:ex:tile-sync}|
    LoadTileToShMem(Ash, A, row, tk); |\label{line:ex:tileA}|
    |\underline{stage.wait({B, col, tk});}| |\label{line:ex:tile-sync-B}|
    LoadTileToShMem(Bsh, B, col, tk); |\label{line:ex:tileB}|
    MultiplyAccumulate(C, Ash, Bsh, 
                        row, col, tk);}|\label{line:ex:mac}|
  |\underline{stage.post({row, col});}| |\label{line:ex:post}||\label{line:ex:kernel-end}|
}
void MLP(int BS, int H, f16* X, f16* W|$_1$|, |\label{line:ex:mlp-start}|
         f16* XW|$_1$|, f16* W|$_2$|, f16* XW|$_{12}$|) {
  dim2 grid1 = {4*H/8, B}/tile1; |\label{line:ex:grid1}|
  dim2 grid2 = {H, B}/tile2; |\label{line:ex:grid2}|
  CuStage<RowMajor, RowSync>|\label{line:ex:alloc-prod}| 
    prod(grid1,tile1);
  CuStage<RowMajor, RowSync>|\label{line:ex:alloc-cons}|
    cons(grid2,tile2);
  // declare prod to cons[XW1] dependency
  CuSync::dependency(prod, cons, XW1); |\label{line:ex:declare-dep}|
  // invoke the producer gemm
  gemm<<<grid1, tb1, prod.stream()>>>
    (X, W|$_1$|, XW|$_1$|, H, prod);|\label{line:ex:mlp-kernel-1}|
  // invoke waitKernel and then consumer 
  cons.waitKernel(); |\label{line:ex:wait-kernel}| 
  gemm<<<grid2, tb2, cons.stream()>>>
    (XW|$_1$|, W|$_2$|, XW|$_{12}$|, 4*H/8, cons);} |\label{line:ex:mlp-kernel-2}|
  \end{lstlisting}
  \caption{
  The kernels are invoked on different streams. 
  The wait kernel ensures the order of kernel invocation. The {\tt post} and {\tt wait} methods ensure tile dependency.
  Changes to the GeMM kernel are \underline{underlined}.
  \label{fig:ex:matmuls}}
  \end{subfigure}
  \hfill{}
  \begin{subfigure}[b]{0.48\textwidth}
  \begin{lstlisting}[language=CUDA, xleftmargin=3.0ex]
class CuStage<Policy>
  void init() {|\label{line:api:cons:init}|
    sems = /*Init semaphores using Policy*/}
  void post(dim2 tile, dim2 grid) {|\label{line:api:prod:post}|
    __syncthreads();  |\label{line:api:prod:post:syncthreads}|
    if(threadIdx == {0,0,0})
    __threadfence_system();|\label{line:api:prod:post:threadfence}|
      sem = &sems[Policy.sem(tile, grid)];
      atomicAdd(sem,1);}|\label{line:api:prod:setval}|
  void wait(dim2 tile, dim2 grid) {|\label{line:api:cons:wait}|
    sem = &sems[Policy.sem(tile, grid)];
    if(threadIdx == {0,0,0})
      while(*sem != Policy.value(tile,grid));|\label{line:api:cons:while}|
    __syncthreads();}|\label{line:api:cons:syncthreads}|

class TileSync|\label{line:api:tilesync-start}|
  int sem(dim2 tile, dim2 grid) {|\label{line:api:tilesync:sem}|
    //Distinct semaphore for each tile
    return tile.x*grid.y + tile.y;}
  int value(dim2 tile, dim2 grid){return 1;|\label{line:api:tilesync:value}|}|\label{line:api:tilesync-end}|

class RowSync|\label{line:api:rowsync-start}|
  int sem(dim2 tile, dim2 grid) {|\label{line:api:rowsync:sem}|
    //Tiles of same row share semaphore
    return tile.y;}
  int value(dim2 tile, dim2 grid) {|\label{line:api:rowsync:value}|
    return grid.x;}|\label{line:api:rowsync-end}|

int RowMajor(dim2 tile, dim2 grid){|\label{line:api:rowmajor}|
    return tile.y*grid.x + tile.x;}
  \end{lstlisting}
  \caption{TileSync creates a semaphore for each tile. 
  RowSync trades concurrency for synchronization by creating a single semaphore per row.\label{fig:ex:tilesync}\label{fig:ex:rowsync}}
  \end{subfigure}
  \caption{Fine-grained synchronization of two GeMMs of MLP using \libname{}'s TileSync and RowSync policies.
  \label{fig:ex}}
  \end{figure*}

  
\subsection{Invoke Dependent Kernels}
\label{sec:invoke-kernel}
The first requirement for fine-grained synchronization is to eliminate the stream synchronization between kernels. 
\libname{} achieves this by invoking all kernels on different CUDA streams. 
The example creates producer (\texttt{prod}) and consumer (\texttt{cons}) stages for both GeMM kernels (lines~\ref{line:ex:alloc-prod}--\ref{line:ex:alloc-cons} in Figure~\ref{fig:ex:matmuls}).
Then, the example declares the dependency between the two stages by specifying that the output of the producer is the input of the consumer (line~\ref{line:ex:declare-dep}). 
Finally, the example invokes both kernels on different streams associated with respective stages (lines~\ref{line:ex:mlp-kernel-1} and~\ref{line:ex:mlp-kernel-2}).
Section~\ref{sec:dependent-tiles} describes how \libname{} enforces this dependency. 

\subsection{Stage Processing Order}
\label{sec:stage-order}
The second requirement for fine-grained synchronization is to execute all full waves of the producer kernel before the consumer kernel.  
However, the CUDA runtime lacks any mechanism to enforce this execution order among kernels belonging to different streams.
Hence, there is a possibility that the consumer kernel is scheduled on the GPU before the producer kernel. 
This can lead to poor performance as thread blocks of the consumer kernel occupy SMs without doing any useful work. 
In the worst case, this can lead to deadlocks if no SMs are available for the producer kernel. 

\libname{} ensures this requirement by enforcing the scheduling order of kernels using its \emph{wait-kernel} mechanism.
The \texttt{wait} kernel is invoked by the consumer stage on the consumer stream before the consumer kernel (line~\ref{line:ex:wait-kernel} in Figure~\ref{fig:ex:matmuls}).
The \texttt{wait} kernel contains a single thread, which waits on a global memory semaphore for each consumer kernel using a busy-wait while loop.
When the producer kernel calls the {\tt stage.start()} function (line~\ref{line:ex:init-prod-cons}), the function sets the semaphore using the first thread of the first thread block, which in-turn exits the \texttt{wait}-kernel.  
After the \texttt{wait}-kernel exits, the CUDA runtime can invoke the consumer kernel.
Thus, \libname{} ensures that no thread blocks of the consumer kernel are scheduled before at least one of the thread blocks of the producer kernel. 

The wait-kernel mechanism assumes that CUDA schedules thread blocks of kernels in the order the kernels are invoked by CUDA.
We have found that the latest versions of CUDA 11 and 12 executing on NVIDIA GPUs based on Volta and Ampere architecture follow this schedule.

\subsection{Custom Tile Order}
\label{sec:tile-order}
The third requirement for efficient synchronization is minimizing waiting time of consumer kernels.
However, the CUDA runtime can schedule thread blocks on SMs in any arbitrary order, which can lead to unpredictable wait times.
Ideally, thread blocks of the consumer kernel should be scheduled in the order the producer kernel generate tiles.

\libname{} enables execution of both producer and consumer kernel's thread blocks in a custom scheduling order independent of how the CUDA runtime schedules thread blocks.
In our example, each thread block calls {\tt stage.tile()} (line~\ref{line:ex:get-tile}) to obtain the tile it needs to compute.
The parameter {\tt RowMajor} (lines~\ref{line:ex:alloc-prod}--\ref{line:ex:alloc-cons}) ensures that both kernels produce tiles in a row major order, i.e., first all thread blocks in \texttt{x}, then in \texttt{y}, and finally in \texttt{z}.
Figure~\ref{fig:ex:rowsync} defines the RowMajor order as a function (line~\ref{line:api:rowmajor}).
A tile order function takes a tile index in the 3-D grid and returns a distinct 1-D index for the tile.
Internally, \libname{} maintains an array that maps a linear tile index to a 3-D index.
For each thread block, \libname{} increments an atomic global counter and returns the 3-D index in the array for the previous counter value.
In summary, \libname{} allows easy experimentation with diverse scheduling orders to obtain the best performance.


\subsection{Synchronizing Dependent Tiles}
\label{sec:dependent-tiles}
The final requirement for fine-grained synchronization is to ensure that the dependence between tiles of producer- and consumer-kernels is maintained using a synchronization mechanism.
\libname{} provides two functions, {\tt wait} and {\tt post} to enforce this dependency. 
For instance, in our example, the {\tt wait} function is called twice (line~\ref{line:ex:tile-sync} and~\ref{line:ex:tile-sync-B}) before loading the tiles of {\tt A} and {\tt B}, and the {\tt post} function is called once (line~\ref{line:ex:post}) for the producer kernel after computing the tile. However, the consumer kernel only needs to wait on the output of the producer kernel, i.e., input \texttt{A} of the consumer kernel.
This dependency is specified in line~\ref{line:ex:declare-dep}.
Therefore, the {\tt wait} before loading a tile of {\tt A} waits for the corresponding {\tt post} of the producer kernel, and the {\tt wait} before loading a tile of {\tt B} becomes a no-op.
Since the producer kernel have no dependency, both {\tt wait}s are no-ops for the producer kernel.

\libname{} provides a mechanism for synchronizing producer and consumer tiles based on an arbitrary \emph{synchronization policy} (or policy in short).
\libname{} uses an array of global memory semaphores for synchronization, where each producer tile is associated with only one semaphore and a semaphore's value represents the status of its producer tiles.
Thus, a policy is a mapping of one or more producer tiles to one semaphore.
For example, the finest grained synchronization policy, we call \emph{TileSync}, waits for each producer tile and is defined as a one-to-one map of a producer tile to a semaphore.
A policy requires implementation of two methods: (i) \texttt{sem}, which returns the semaphore for the given tile, and (ii) \texttt{value}, which returns the expected value of semaphore when the tile is ready.
We below describe details of three methods of \texttt{CuStage} required for our synchronization mechanism (lines~\ref{line:api:cons:init}--\ref{line:api:prod:setval} in Figure~\ref{fig:ex:rowsync}).

\spara{init}: The \texttt{init} method allocates and initializes the array of semaphores in the global memory based on the given policy.
 
\spara{post}: The \texttt{post} method calls \texttt{\_\_syncthreads} and a memory fence to ensure that all threads of the thread block has computed the tile and all global memory writes are visible to other kernels (line~\ref{line:api:prod:post:syncthreads}--\ref{line:api:prod:post:threadfence}).
Finally, the method obtains the semaphore for the tile using the policy and increments the semaphore (line~\ref{line:api:prod:setval}).

\spara{wait}: The \texttt{wait} method obtains the semaphore for the given tile using the policy and then wait on the value of semaphore in a while loop using only the first thread of the thread block (line~\ref{line:api:cons:while}).
While the first thread is waiting, all other threads of the thread block are blocked on the \texttt{\_\_syncthreads} (line~\ref{line:api:cons:syncthreads}).
When the semaphore changes to the expected value, all threads of the thread-block proceeds from the \texttt{\_\_syncthreads}.

\subsection{Synchronization Policies}
\label{sec:synchronization-policies}
\libname{} allows implementation of diverse synchronization policies easily.
As described earlier, each policy requires implementing \texttt{sem} and \texttt{value} methods.
Below we discuss two general policies that are applicable to all kernels in our workloads.

\spara{TileSync} is the finest-grained policy that synchronizes on each producer tile (lines~\ref{line:api:tilesync-start}--\ref{line:api:tilesync-end} in Figure~\ref{fig:ex:tilesync}).
To minimize the wait time of the consumer-kernel, both kernels compute their tiles in a row major order.
The \texttt{sem} method returns distinct semaphore for each tile (line~\ref{line:api:tilesync:sem}) and the \texttt{value} method returns 1 to signify that the tile is computed (line~\ref{line:api:tilesync:value}).
For example, in Figure~\ref{fig:ex:matmuls} to compute a tile $E^{xy}$, the TileSync policy requires waiting first on $C^{x0}$ and then on $C^{x1}$.

\spara{RowSync} synchronizes on each row of the producer kernel requiring less synchronizations than TileSync (lines~\ref{line:api:rowsync-start}--\ref{line:api:rowsync-end} in Figure~\ref{fig:ex:tilesync}).
For example, for two GeMMs of Figure~\ref{fig:ex:matmuls}, TileSync requires 12 synchronizations in total,
while RowSync requires 6 synchronizations by sharing the same semaphore for all tiles computing the same row of $C$.
Thus, the \texttt{sem} method returns the row of the given tile and the \texttt{value} method returns the value when the row is ready, i.e., the number of tiles in a row (line~\ref{line:api:rowsync:sem}--\ref{line:api:rowsync:value}).
To minimize the wait time, both kernels schedule their tiles in a row major order.
RowSync can also be used for synchronizing Conv2D kernels. 
Section~\ref{sec:eval} shows that for large GeMMs and Conv2Ds the high number of synchronizations is a bottleneck.

\begin{figure}
\small
\begin{subfigure}{\columnwidth}
\begin{lstlisting}[language=CUDA, xleftmargin=3.0ex]
Dim x, y;|\label{line:dep:dims}|
//Max value of all dimensions of both GeMMs
Grid g1(x, y, $\frac{\mathtt{H}}{2*\mathtt{TileN}}$, $\frac{\mathtt{B*S}}{\mathtt{TileM}}$);
Grid g2(x, y, $\frac{\mathtt{H}}{\mathtt{TileN}}$, $\frac{\mathtt{B*S}}{\mathtt{TileM}}$);|\label{line:dep:grids}|
//Tile is produced by each thread block
Tile prod(x, y), cons(x, y);|\label{line:dep:tiles}|
//All col tiles for a row from 0 to $\textcolor{commentgreen}{\frac{\mathtt{H}}{2*\mathtt{TileN}}}$
ForAll prodCols(prod, x, Range(g1.x));|\label{line:dep:alltiles}|
//Tile of 2nd GeMM depends on all
//col tiles of 1st GeMM
Dep dep({g2, cons}, {g1, prodCols});|\label{line:dep:dep}|
\end{lstlisting}
\caption{GPT-3's MLP\label{fig:dependency-dag}}
\end{subfigure}
\begin{subfigure}{\columnwidth}
\begin{lstlisting}[language=CUDA, xleftmargin=3.0ex]
Dim x, y;
//First GeMM Grid
Grid g1(x, y, $\frac{\mathtt{3*H}}{8\mathtt{*TileN}}$, $\mathtt{\frac{B*S}{TileM}}$);
//P, R, and T Grid
Grid gP(x, y, $\frac{\mathtt{B*(S+S')}}{\mathtt{TileN}}$, $\frac{\mathtt{B*(S+S')}}{\mathtt{TileM}}$);
Grid gR(x, y, $\frac{\mathtt{B*(S+S')}}{\mathtt{TileN}}$, 1);
Grid gT(x, y, $\frac{\mathtt{B*(S+S')}}{\mathtt{TileN}}$, $\frac{\mathtt{H}}{8\mathtt{*TileM}}$);
//Second GeMM Grid
Grid g2(x, y, $\frac{\mathtt{H}}{8\mathtt{*TileN}}$, $\mathtt{\frac{B}{TileM}}$);
//P to 1st GeMM
//Strided Tile Dependencies: stride=$\color{commentgreen}{\frac{\mathtt{H}}{8\mathtt{*TileN}}}$
Dep dep1P({gP, Tile(x,y)}, 
  {g1, Tile(x,y), Tile(x+$\frac{\mathtt{H}}{8\mathtt{*TileN}}$,y)})|\label{line:Attention:dep12}|;
Dep depPR({gR, Tile(x,y)},
  {gP, ForAll(Tile(x,y), y, Range(gP.y))});
Dep depTR1({gT, Tile(x,y)}, 
  {gR, Tile(x, y)}, {g1, Tile(x+$\frac{2*\mathtt{H}}{8\mathtt{*TileN}}$,y)});
//2nd GeMM to T
dep23({g3, Tile(x,y)}, {gT, Tile($\mathtt{\frac{x}{TileM}}$,y)})|\label{line:Attention:dep23}|;
\end{lstlisting}
\caption{Attention\label{fig:use-case:Attention}}
\end{subfigure}
\begin{subfigure}{\columnwidth}
\begin{lstlisting}[language=CUDA, xleftmargin=3.0ex]
Dim x, y;
//First GeMM Grid
Grid g1(x, y, $\mathtt{\frac{C}{TileM}}$, $\mathtt{\frac{B*P*Q}{TileN}}$);
//Second GeMM Grid
Grid g2(x, y, $\mathtt{\frac{C}{TileM}}$, $\mathtt{\frac{B*P*Q}{TileN}}$);
//2nd Conv2D to 1st Conv2D
Dep dep({g2, Tile(x,y)}, {g1, Tile($\mathtt{\frac{x}{R*S}}$,y)});|\label{line:use-case:conv2d:dep}|
\end{lstlisting}
\caption{Two Conv2Ds\label{fig:use-case:conv2d}}
\end{subfigure}
\caption{Dependencies in the \compiler{} DSL. 
\texttt{TileM} and \texttt{TileN} are tile size of GeMMs in row and column respectively.}
\end{figure}


\section{Auto-tuning of Policies and Tile Orders}
\label{sec:cusyncgen}
The process of obtaining the best performance involves experimenting with several synchronization policies and tile processing orders.
The best policy and tile order depends on computations, data sizes, and the GPU architecture.
However, doing this process manually is both tedious and error-prone.

Therefore, \compiler{} is a tool that takes dependencies specified by the user and generates the optimal tile processing order and multiple synchronization policies as CUDA code for \libname{}.
\libname{} currently requires the user to manually modify the GPU kernels to instantiate \texttt{CuStage} with generated policies and tile processing order similar to the MLP example (Figure~\ref{fig:ex:matmuls}).
The modularity of \libname{} allows the user to easily plug diverse synchronization policies and tile processing orders.
\subsection{Workflow}
The workflow of \compiler{} is as follows:
\begin{enumerate}[leftmargin=4mm]
  \item The user describes a chain of dependencies between kernel tiles and the grid values for all kernels.
  \item \compiler{} checks bounds of producer and consumer tiles based on grid values.
  \item \compiler{} generates a tile processing order as CUDA code that minimizes the wait time.
  \item \compiler{} generates CUDA code for multiple policies.
  \item The user modifies the workload to support \libname{} and plugs the generated CUDA code to \libname{}.
\end{enumerate}
The rest of the section describes each of these steps.

\spara{Describe Dependencies}
The user describes dependencies between tiles of kernels using a DSL embedded in C++.
Figure~\ref{fig:dependency-dag} shows the dependency between both GeMMs of MLP described in the DSL.
First, the DSL code must define each kernel's grid dimensions with their maximum value.
The example defines \texttt{x} and \texttt{y} dimensions for both grids (line~\ref{line:dep:dims}--\ref{line:dep:grids}).
Specifying the exact values for a grid enables generating efficient code and doing bounds checking for correctness.
Then, the DSL code constructs producer and consumer tiles by specifying an affine function over each dimension of the grid.
The example creates a producer and consumer tile for each thread block in the grid and creates a range of column tiles using \texttt{ForAll} (line~\ref{line:dep:tiles}--\ref{line:dep:alltiles}).
Finally, the code specifies the dependence between one consumer tile and one or more producer tiles (line~\ref{line:dep:dep}).

\spara{Generate Tile Processing Order}
\compiler{} generates a tile processing order for each kernel to minimize the waiting time.
To discuss the process, consider a dependency where a consumer tile, $C(x, y)$, depends on $N$ producer tiles, \{$P(x, a_0y+b_0), P(x, a_1y+b_1), \ldots, P(x, a_{N-1}y+b_{N-1})$\}.
We achieve minimum wait time when the consumer kernel consumes tiles in the same order as they are produced by the producer kernel.
Thus, we schedule all $N$ producer tiles consecutively for each consumer tile using the following code:

\begin{lstlisting}[language=CUDA, xleftmargin=3.0ex]
int prodOrder(dim2 tile, dim2 grid) {
  int linear = bid.y*gDim.x + bid.x, y = 0;
  if (tile.y%$\mathtt{a_0}$ <= $\mathtt{b_0}$) y = 0;
  //Similarly for tiles till N-2
  else if (tile.y%$\mathtt{a_{N-1}}$ <= $\mathtt{b_{N-1}}$) y = N-1;
  return linear/N+y;}
\end{lstlisting}

This code obtains the 1-D linear index of a tile, finds the tile index within the group of $N$ tiles, and returns the new linear index.
We also set the consumer kernel to follow the row major order of tiles. 
Our MLP example uses the row major order, i.e., all groups of $\frac{\mathtt{H}}{\mathtt{TileN}}$ consecutive producer tiles are scheduled consecutively.
It is straightforward to extend this method to a chain of dependent kernels by extending the dependence from the last consumer kernel to the first producer kernel and then generating code for each kernel.

\spara{Generating Policies}
\compiler{} generates multiple synchronization policies for each dependence.
For the following discussion, consider a dependence where a consumer tile, $C(x, y)$, depends on $N$ producer tiles, \{$P(x, a_0y+b_0), P(x, a_1y+b_1), \ldots, P(x, a_{N-1}y+b_{N-1})$\}.
\compiler{} generates two policies for the dependence in each dimension:
(i) map each tile to a distinct semaphore, or (ii) map all $N$ tiles to the same semaphore. 
The code generated for the considered dependence and the value of $M \in \{1, N\}$ is:

\begin{lstlisting}[language=CUDA, xleftmargin=3.0ex]
int sem(dim2 tile, dim2 grid) {
  int y = 0;
  if (tile.y%$a_0$ <= $b_0$) 
    y = (tile.y-$b_0$)/$a_0$;
  //Similarly for tiles till M-2
  else if (tile.y%$a_{M-1}$ <= $b_{M-1}$) 
    y = (tile.y-$b_{M-1}$)/$a_{M-1}$;
  else y = tile.y;
  return y*grid.x + tile.x;}
int value(dim2 tile, dim2 grid) {return M;}
\end{lstlisting}

After considering both cases for the innermost dimension, the phase moves to the outer dimension, and follows the same method.
In our MLP example, \compiler{} generates two policies: (i) TileSync that maps each tile to a distinct semaphore, and (ii) RowSync that maps all column tiles of the same row to the same semaphore.

\spara{Running the Generated Code}
We require the user to modify the workload to support running \libname{} by adding \texttt{wait} calls before every tile load and \texttt{post} call after computing a tile.
For example, in the case of MLP, we require the user to do the changes of Figure~\ref{fig:ex:matmuls}.
The modularity of \libname{} enables plugging multiple policies and tile processing order.
So, the user can execute all generated policies and obtain the policy with least execution time.

\subsection{Computation Dependencies in ML Models}
We now show how to specify dependencies of Attention and Conv2D cases in \compiler{}.

\spara{Attention} contains two dependencies between its three kernels (Figure~\ref{fig:use-case:Attention}).
In the first dependency, an element of the dot product depends on three elements in the same row with a stride of $\mathtt{\frac{H}{8}}$ of the first GeMM output (line~\ref{line:Attention:dep12}).
In addition to TileSync and RowSync, for this dependence \compiler{} also generates a policy, we call \emph{StridedSync}, that maps all three producer tiles of the first GeMM to the same semaphore.
Thus, \emph{StridedSync} waits until all three tiles of the first GeMM are computed before continuing with the dot product of tiles. 
Moreover, \compiler{} generates the tile order that schedules these three tiles consecutively.
For other dependencies, \compiler{} generates both TileSync and RowSync, while processing tiles in a RowMajor order.


\spara{Conv2D} using the implict GeMM algorithm converts a convolution of \texttt{B} input images of size $\threeDShape{P}{Q}{C}$ with a kernel matrix of size $\twoDShape{R}{S}$ into a GeMM of matrices $\twoDShape{B*P*Q}{C*R*S}$ with $\twoDShape{C*R*S}{C}$.
Figure~\ref{fig:use-case:conv2d} shows the dependency between two Conv2Ds using the implicit GeMM algorithm.
Thus, the dependency describes that each tile of the second implicit GeMM depends on all column tiles of the first implicit GeMM output (line~\ref{line:use-case:conv2d:dep}).
\compiler{} generates two policies for this dependency: (i) RowSync to synchronize each row, and (ii) \emph{Conv2DTileSync} policy to synchronize each tile.
Moreover, \compiler{} generates a row major order for both Conv2Ds.

\subsection{Optimizations}
\label{sec:opt}
\compiler{} automatically perform several optimizations to improve the performance of a \libname{} synchronized workload.
These optimizations depend on the architecture details of the GPU, occupancy of CUDA kernels, and grid sizes.
The optimizations are as follows:

\spara{Avoid Wait Kernel}
The wait-kernel mechanism ensures that all thread blocks of the producer kernel are scheduled on the GPU before the consumer kernel.
However, if both producer and consumer kernels can be executed in less than two waves, we do not need the wait-kernel mechanism.

\spara{Avoid Custom Tile Processing Order}
We can also avoid a custom tile processing order when all tiles of producer and consumer-kernels can be executed in two waves.

\spara{Reorder Tile Loads and Synchronization} \label{sec:opt:reorder}
The general workflow of tile based CUDA kernels is to load a tile of all inputs into shared memory or registers and then perform operations on all tiles.
We can re-order the waiting of tile of one input with the loading of other tile, to overlap the waiting of one tile with the loading of the other input's tile.
For example, in Figure~\ref{fig:ex:matmuls} the second GeMM kernel loads a tile of both inputs (\texttt{A} and \texttt{B}) and compute the tile of output matrix (\texttt{C}) (line~\ref{line:ex:tile-sync}--\ref{line:ex:tileB}).
We can reorder the loading of \texttt{B} tile with the waiting on \texttt{A} tile, i.e., swap lines~\ref{line:ex:tile-sync}--\ref{line:ex:tileA} with lines~\ref{line:ex:tile-sync-B}--\ref{line:ex:tileB}.
Since there is no waiting for tile of \texttt{B}, loading a \texttt{B} tile can overlap with waiting of \texttt{A} tile, leading to improved performance.
\compiler{} automatically performs the reordering if the user annotate tile loading in kernels with \texttt{\#pragma tile}.

\section{Evaluation}
\label{sec:eval}
We now evaluate the performance of \libname{} against state-of-the-art baselines using large open source ML models as our workloads.

\subsection{Experimental Setup}
We run our experiments on a machine with a 2.60GHz 12-core Intel Xeon E5-2690 CPU with 448GB RAM and 8 NVIDIA Tesla V100 32GB GPUs connected with NVLINK.
We use CUDA 12.2 and report the average time of 20 executions after a warmup of 5 executions.

\spara{ML Models} We used \libname{} to synchronize CUDA kernels of four ML models: MegatronLM GPT-3 145 Billion~\cite{megatronlm}, LLaMA 65.2 Billion~\cite{touvron2023llama}, ResNet-38~\cite{resnet}, and VGG-19~\cite{vgg}.
We used the GeMM and Conv2D CUDA kernels of NVIDIA CUTLASS 3.1 (Figure~\ref{fig:gpt-3:attention}).
We fuse the pointwise computations with GeMM and Conv2D kernels and developed a fused kernel of Softmax and Dropout in the Attention.
We evaluate the reduction in inference times of these models using \libname{} on batch sizes from 1 to the largest supported batch size by each model.

\spara{Baselines} We consider the following baselines:\\
\textbf{StreamSync} is the CUDA stream synchronization.\\
\textbf{Stream-K}~\cite{streamk} partitions the last thread block wave of GeMM and Conv2D among all SMs to improve the GPU utilization.

\begin{table}[t]
  \vspace{1em}
  \centering
  \caption{\sc Fraction of lines of code changed in GeMM, Fused Softmax-Dropout, and Conv2D kernels to support using \libname{}.\label{tab:loc}}
  \begin{tabular}{|r|r|r|r|r|}
    \hline
    \textbf{Kernel} & \textbf{Implementation} & \multicolumn{2}{c|}{\textbf{Lines Changed}}\\
    \cline{3-4}   
          &         & \textbf{Number} & \textbf{Fraction}\\\hline 
    GeMM & CUTLASS & 25& 0.5\%\\
    Softmax-Dropout & Ours & 5& 1\%\\
    Conv2D & CUTLASS & 22& 0.6\%\\
    \hline
  \end{tabular}
\end{table}

\begin{figure*}[th]
  \begin{subfigure}[b]{0.26\textwidth}
    \includegraphics[scale=0.65]{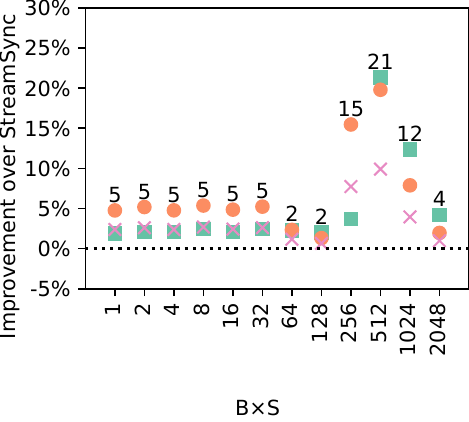}
    \caption{GPT-3 MLP\label{fig:results:gpt-3:mlp}}
  \end{subfigure}
  \hfill{}
  \begin{subfigure}[b]{0.23\textwidth}
    \includegraphics[scale=0.65]{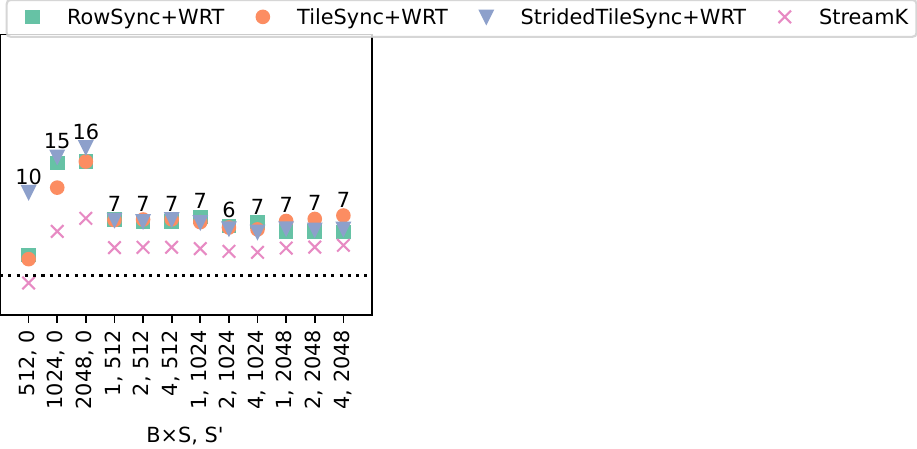}
    \caption{GPT-3 Attention\label{fig:results:gpt-3:attention}}
  \end{subfigure}
  \begin{subfigure}[b]{0.23\textwidth}
    \includegraphics[scale=0.65]{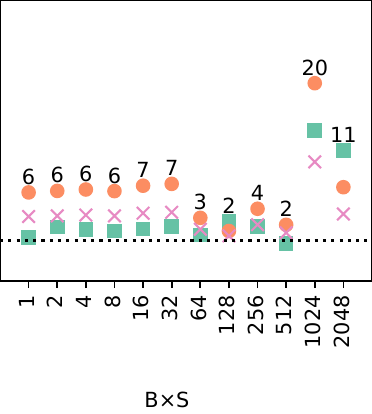}
    \caption{LLaMA MLP\label{fig:results:llama:mlp}}
  \end{subfigure}
  \begin{subfigure}[b]{0.23\textwidth}
    \includegraphics[scale=0.65]{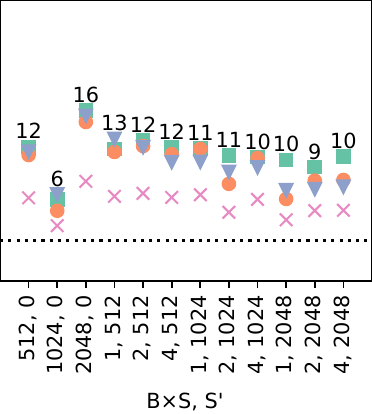}
    \caption{LLaMA Attention\label{fig:results:llama:attention}}
  \end{subfigure}
  \caption{Improvement of \libname{}'s policies and StreamK in MLP and Attention over StreamSync for batch sizes 1--2048.
  During prompt processing $\texttt{S'} = 0, \texttt{B}\times \texttt{S} > 1$ and in token generation $\texttt{S'} > 1, \texttt{B} \geq 1, \texttt{S} = 1$.
  Numbers shows the maximum speedup out of all policies.\label{fig:results:gpt-3}}
\end{figure*}

\begin{table*}
  \centering
  \caption{\sc Grid size, number of waves, total waves and execution time in StreamSync and \libname{} for both GeMMs of GPT-3's MLP.
  The grid \texttt{x} and \texttt{y}-dims are obtained by dividing the size of GeMM with the tile size and the \texttt{z}-dim is the number of thread blocks used for split-k.\label{tab:waves:mlp}}
  \begin{tabular}{|r|r|r|r|r|r|r|r|r|r|r|}
  \hline
  \textbf{Batch}  & \multicolumn{2}{c|}{\textbf{First GeMM}} & \multicolumn{2}{c|}{\textbf{Second GeMM}} & \multicolumn{2}{c|}{\textbf{StreamSync}} & \multicolumn{3}{c|}{\textbf{\libname{}}} & \textbf{Decrease in}\\
    \cline{2-3} \cline{4-5} \cline{6-7} \cline{8-10}
    \textbf{Size} & \multicolumn{1}{c|}{\textbf{Grid}} & \textbf{Waves} & \multicolumn{1}{c|}{\textbf{Grid}} & \textbf{Waves} & \textbf{Waves} & \textbf{Time($\mu$s)} &\textbf{Waves} & \textbf{Policy} & \textbf{Time($\mu$s)} & \textbf{Runtime}\\ \hline
    1--64 & 1$\times$24$\times$4&0.6&1$\times$48$\times$3 &0.9& 2 & 378& 1.8 & Tile & 355&5--6.0\%\\
    128   & 1$\times$24$\times$3&0.4&1$\times$48$\times$3 &0.9& 2 & 530& 1.3 & Tile & 523&2\%\\
    256   & 1$\times$48$\times$4&1.2&1$\times$96$\times$2 &1.2& 4 & 862& 2.4 & Tile & 728&16\%\\
    512   & 2$\times$24$\times$2&1.2&2$\times$48$\times$1 &1.2& 6 & 1500& 4.8 & Row  & 1196&21\%\\
    1024  & 4$\times$24$\times$2&2.4&4$\times$48$\times$1 &2.4& 5 & 2111& 3.6 & Row  & 1901&10\%\\
    2048  & 8$\times$24$\times$1&2.4&8$\times$48$\times$1 &4.8& 8 & 3730& 7.2 & Row  & 3574&4\%\\
    \hline
  \end{tabular}
\end{table*}


\subsection{Ease of Programming}
Table~\ref{tab:loc} shows that the number of lines added and changed to support fine-grained synchronization of GeMM, Conv2D, and Softmax-Dropout kernels using \libname{} are negligible compared to the lines of code of these kernels.
Thus, the \libname{} approach enables diverse synchronizations of tile based computation kernels through few modifications.

\subsection{Applicability in ML Models Inference}
We now discuss the applicability of \libname{} in improving the performance of ML models from the perspective of kind of computations and the average utilization of GPU.
First, ML models majorly consists of tile based GPU kernels, such as GeMM and Conv2Ds.
Since \libname{} supports any tile based kernel, we can use \libname{} to synchronize kernels of ML models.
Second, since the number of waves of each kernel increases with the batch size, the average utilization of all waves also increases.
However, each ML model supports a maximum batch size limit during both training and inference phases.
For example, the maximum token length supported by GPT-3 and LLaMA is 2048.
We show in our experiments that even for this maximum batch size, GPU kernels suffer from low number of waves leading to low average utilization.
In summary, \libname{} is applicable to diverse ML models because ML models largely contains tile based kernels and the maximum batch size supported by ML models still suffers from under-utilization.

\subsection{Maximum Overhead of Synchronization}
\change{
The synchronization mechanism has two sources of overhead: global memory accesses and \texttt{\_\_syncthreads}.
The percentage of total overhead depends on the amount of computations performed by the GPU kernel.
A kernel doing large amount of computations on each tile would suffer from less synchronization overhead than a kernel doing less amount of computations.
We can obtain an upper bound on the overhead by having two kernels (i) doing minimum computations on each tile, (ii) execute maximum number of thread blocks per wave, and (iii) execute one full wave.

We design such an experiment where the producer kernel copies values from an input array to an intermediate output array by assigning consecutive threads to contiguous elements, and similarly the consumer kernel copies values from the intermediate array to a final output array.
Thus, a thread block of the consumer depends on the same thread block of the producer.
We invoke both kernels with the maximum number of thread blocks per wave on Tesla V100, i.e., $Number\_of\_SMs \times Max\_Occupancy=80\times 16 = 1280$.
We found that synchronization using \libname{} leads to 2-3\% overhead over StreamSync.
Hence, \libname{}'s synchronization mechanism provides low overhead.
}

\subsection{Large Language Model Inference Results}
We now evaluate the reduction in the inference times of GPT-3 and LLaMA with model parallelism on 8 GPUs using \libname{} for both prompt processing and token generation phase (Figure~\ref{fig:gpt-3}).
In prompt processing, we consider the total number of tokens in an inference task, i.e., \texttt{B$\times$S} from 512 to 2048, and in token generation, we consider batched requests, i.e., \texttt{B} from 1 to 4 with number of already generated tokens, i.e. \texttt{S'} from 512 to 2048.
We used \compiler{} to generate the following policies:\\
\textbf{RowSync+WRT} synchronizes rows and executes thread blocks in the row major order by adding our optimizations of Section~\ref{sec:opt}, i.e., avoiding the wait-kernel (W), avoiding custom tile order (T), and reorder tile loads (R).\\
\textbf{TileSync} synchronizes tiles and executes thread blocks in the row major order.\\
\textbf{TileSync+WRT} extends TileSync by adding our optimizations of Section~\ref{sec:opt}.\\
\textbf{Strided+TileSync+WRT}, only for Attention, synchronizes the first GeMM with the first GeMM of Cached mechanism using StridedSync, and all other kernels using TileSync (Figure~\ref{fig:use-case:Attention}).
The policy also add our optimizations of Section~\ref{sec:opt}.

\subsubsection{MLP Results}
Figure~\ref{fig:results:gpt-3:mlp} and \ref{fig:results:llama:mlp} shows that synchronizing dependent GeMMs of the GPT-3 MLP and LLaMA MLP using \libname{} decreases the execution time of both MLPs by up to 20\% for different sizes.
We discuss these results using Table~\ref{tab:waves:mlp} that shows the number of waves for all batch sizes for GPT-3 MLP using both StreamSync and \libname{}.

TileSync+WRT performs best for \texttt{B$\times$S} of 1 to 256 because there is a single thread block in the \texttt{x}-dimension of grid (Table~\ref{tab:waves:mlp}). 
The improvement at size 256 is higher than small sizes because 
TileSync+WRT reduces the number of waves by 1 over StreamSync.
On small batch sizes, even though the number of waves is not decreased, TileSync+WRT performs 7\% faster because the second GeMM can overlap the loading of \texttt{W$_2$} tile into the shared memory with the computation of the first GeMM.

RowSync performs best for sizes greater than 512 because synchronizing over a row once reduces memory accesses than synchronizing over multiple tiles and more number of rows provides more opportunities for overlapping.
Therefore, increasing the number of rows also increases the speedup of RowSync from 4\% at 256 to 20\% at 1024.
However, the speedup decreases to 4\% at 2048 because the fraction of waves reduced by \libname{} decreases with more thread blocks in the grid.

\spara{Effect of Overlapping Kernel Invocations} We measured the time of a kernel invocation is $\approx$6$\mu$s, which is significantly lower than the difference in the execution time of StreamSync and \libname{}.
Table~\ref{tab:waves:mlp} shows that the difference in execution times with \libname{} and StreamSync is significantly higher than the time to invoke a kernel.
Hence, the performance improvement of \libname{} is significantly higher than what would be achieved by only overlapping the invocation of the second GeMM with the first GeMM execution.

\subsubsection{Attention Results}
Figure~\ref{fig:results:gpt-3:attention} shows that synchronizing all kernels of Attention using \libname{} provides 6-16\% improvement over StreamSync for both GPT-3 and LLaMA.

During prompt processing, i.e. when \texttt{S$'$ = 0}, StridedTileSync+WRT works better than both RowSync+WRT and TileSync+WRT because StridedTileSync+WRT performs less number of synchronizations than TileSync and provides larger overlapping opportunities than RowSync. 
During token generation, i.e. when \texttt{S$'$ = 1} and \texttt{S = 1}, all policies works similarly because different synchronization policies provides best performance between different kernels.

\begin{figure*}[th]
  \begin{subfigure}[b]{0.4\textwidth}  
    \includegraphics[scale=0.63]{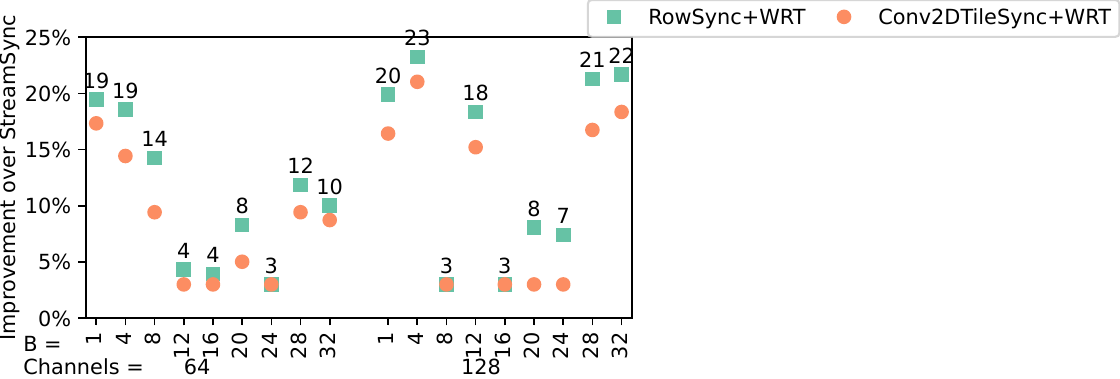}
    \caption{2x Conv2Ds per layer in ResNet-38 and \change{
      VGG-19}\label{fig:results:conv}}
  \end{subfigure}
  \begin{subfigure}[b]{0.29\textwidth}  
    \includegraphics[scale=0.63]{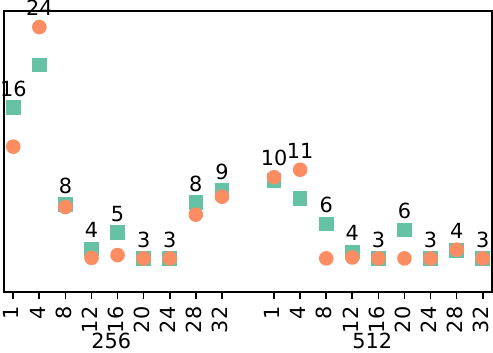}
    \caption{2x Conv2Ds per layer in ResNet-38\label{fig:results:conv}}
  \end{subfigure}
  \hfill{}
  \begin{subfigure}[b]{0.29\textwidth}  
    \includegraphics[scale=0.63]{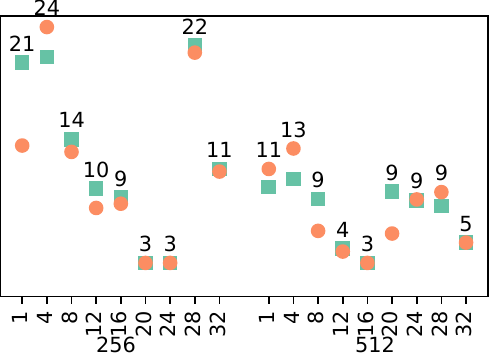}
    \caption{\change{
      4x Conv2Ds per layer in VGG-19\label{fig:results:vgg}}}
  \end{subfigure}
  \caption{
    Performance improvement of \libname{} policies for all Conv2D kernels of each layer over StreamSync in ResNet-38 and \change{
      VGG-19} for different batch sizes.\label{fig:results:resnet}}
\end{figure*}

\subsection{Computer Vision Model Inference Results}
We now evaluate the decrease in inference times of Resnet-38 and \change{VGG-19} by synchronizing all Conv2D kernels of each layer of both models using \libname{} (Table~\ref{tab:resnet-sizes}).
We used \compiler{} to generate the following policies:\\
\textbf{RowSync+WRT} synchronizes rows and execute thread blocks in a row major order with our optimizations in Section~\ref{sec:opt}, i.e., apply avoid wait-kernel (W), avoiding custom tile ordering (T), and reordering tile loads optimizations (R). \\
\textbf{Conv2DTileSync} synchronizes tiles of Conv2Ds and execute thread blocks in a row major order.\\
\textbf{Conv2DTileSync+WRT} extends Conv2DTileSync with our optimizations in Section~\ref{sec:opt}.

Figure~\ref{fig:results:conv} shows that synchronizing all Conv2D kernels of each layer of ResNet-38 and VGG-19 using \libname{} provides up to 24\% improvement over StreamSync for different channels and batch sizes.
For each channel, the improvement follows an oscillating behavior with increasing batch size, i.e., increases to a local maximum then decreases to a local minimum and finally increases to another local maximum.
For example, for 128 channels, the improvement increases from 20\% at batch size 1 to 24\% at batch 4 and then decreases to 3\% at batch size 8, while increasing again to 18\% at batch size 12 and then again decreases to 3\% at batch size 16.
This oscillating behavior is due to the fact that increasing batch size increases invoked number of thread blocks leading to the oscillating behavior of fraction of waves reduced by \libname{}.


\begin{figure}[th]
  \begin{subfigure}[b]{0.55\columnwidth}
  \includegraphics[scale=0.67]{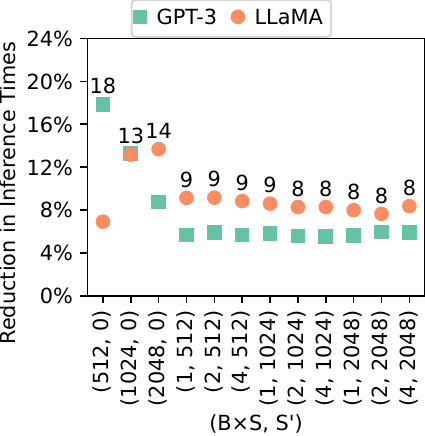}
  \caption{Language Models}
  \end{subfigure}
  \begin{subfigure}[b]{0.42\columnwidth}
  \includegraphics[scale=0.67]{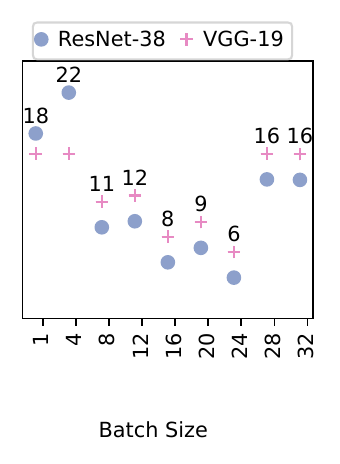}  
  \caption{Vision Models}
  \end{subfigure}
  \caption{\label{fig:results:e2e}Reduction in end-to-end inference times of using \libname{}.}
\end{figure}

\subsection{End-to-End Inference of ML Models}
We integrated \libname{} synchronized CUDA kernels in all four ML models and then evaluate the improvement in end-to-end inference times of these models.
Figure~\ref{fig:results:e2e} shows that using \libname{} synchronized kernels decreases the inference times of GPT-3 by 6--15\%, LLaMA by 9--13\%, 
of ResNet-38 by 5--22\%, and VGG-19 by 6--16\%.
Hence, \libname{} significantly reduces the inference times of popular ML models.

\subsection{Comparison with Stream-K}
We also evaluated the performance of \libname{} against Stream-K for GeMMs kernels.
The best policy of \libname{} performs up to 15\% better than Stream-K in GPT-3 and LLaMA (Figure~\ref{fig:results:gpt-3}).
The speedup of \libname{} over Stream-K is because Stream-K divides the GeMM workload into two kernel calls.
The first kernel computes GeMM using the traditional tiled approach for full waves while the second kernel partitions workload of the final wave among all SMs.
This design requires multiple memory accesses while \libname{} performs a single atomic add to post the status of a computed tile and a read to wait on the status of a producer tile.
Moreover, it is not straightforward to apply the idea of Stream-K to all tile-based kernels.
Currently, Stream-K only supports GeMM computations in NVIDIA CUTLASS.
This is why we cannot apply Stream-K to Conv2D, while \libname{} is valid for any tile based kernels.

\subsection{Impact of Optimizations}
We now discuss the performance improvements provided by the optimizations on top of TileSync for ResNet-38 and GPT-3's MLP.
Table~\ref{tab:impact-opt:mlp} shows that applying all optimizations decreases execution times for kernels with low thread blocks.

\begin{table}[t]
  \vspace{1em}
  \centering
  \caption{\sc Execution times of TileSync with optimizations in GPT-3's MLP and Conv2DTileSync in ResNet for smaller grid sizes. \emph{+W} avoids the wait-kernel. \emph{+WR} also reorders the tile loading. \emph{+WRT} also avoids custom tile order.\label{tab:opt-impact}}
  \begin{subtable}[t]{\columnwidth}
    \centering
    \caption{\sc Execution times in $\mu s$ of TileSync of GeMM kernels in GPT-3's MLP with and without optimizations for different batch sizes.\label{tab:impact-opt:mlp}}
    \begin{tabular}{|r|r|r|r|r|r|r|r|r|}
      \hline
      \textbf{B} & \multicolumn{4}{c|}{\textbf{TileSync}}\\
      \cline{2-5}
        & \textbf{Vanilla} & \textbf{+R} & \textbf{+WR} & \textbf{+WRT}\\\hline
       1--64 & 378 &  365 & 360 & 355\\ \hline
     \end{tabular}
  \end{subtable}

  \begin{subtable}[t]{\columnwidth}
    \caption{\sc Execution times $\mu s$ of Conv2DTileSync of ResNet-38's Conv2D with and without optimizations for all channels and small batch sizes.\label{tab:impact-opt:resnet}}
    \centering
  \begin{tabular}{|r|r|r|r|r|r|r|r|r|r|}
    \hline
    \textbf{C} & \textbf{B} & \multicolumn{4}{c|}{\textbf{Conv2DTileSync}}\\
    \cline{3-6}
    & & \textbf{Vanilla} & \textbf{+R} & \textbf{+WR} & \textbf{+WRT}\\\hline
    64       & 1  & 50 & 45   & 41 & 37\\
    128      & 1  & 60 & 56   & 50 & 45\\
    256      & 1  & 65  & 61   & 56 & 51\\ \hline
\multirow{2}{*}{512} & 1 &100 &94    & 89 & 85\\ 
                     & 4 &135&128   & 120 &115\\\hline
  \end{tabular}
\end{subtable}

\end{table}

\section{Related Work}
Several works have focussed on efficient software-based synchronization between threads of the same CUDA kernel for irregular applications~\cite{10.1145/2751205.2751232, 10.1145/2903150.2903155, 10.1145/3297858.3304055}.
Li et. al.~\cite{10.1145/2751205.2751232} developed an approach for inter-thread synchronizations by reassembling the micro-instructions of shared memory atomic operations in an efficient manner.
Kai et. al.~\cite{10.1145/3297858.3304055} presented a hierarchical synchronization approach  for irregular applications by synchronizing thread blocks using global memory and threads of a thread block using shared memory.
Xu et. al.~\cite{10.1145/2903150.2903155} present a lock design
that uses lock stealing to avoid deadlocks.
\textsc{CoCoNet}~\cite{10.1145/3503222.3507778} performs synchronization between computation and communication kernel to overlap the communication transfers with the computation.
\libname{} targets synchronization between threads of multiple CUDA kernels and provide abstraction to easily design several synchronization policies, both of these are missing from above mentioned works.
\balance
Moreover, some works have focussed on hardware-supported synchronization primitives for inter-kernel threads.
GLocks~\cite{10.1109/IPDPS.2011.87} is the first hardware supported implementation for highly-contented locks using message passing.
HQL~\cite{10.1109/IPDPS.2013.82} is a hardware-accelerated fine-grained lock scheme for GPUs, which adds support for queuing locks in L1 and L2 caches and uses a customized communication protocol for faster lock transfer and reduced lock retries.
ElTantway et. al.~\cite{8327023} propose a hardware warp scheduling policy that reduces lock retries by de-prioritizing warps whose threads are waiting in their spin lock.
They also propose a hardware mechanism for accurately detecting busy-wait synchronization on GPUs.
Dalmia et. al.~\cite{9933620} designed multi-level barrier and priority mechanisms for semaphores for GPU based synchronization primitives. 
\libname{} is a software solution for synchronizing threads of multiple CUDA kernels and these hardware-supported mechanisms are complementary to \libname{}.

Lingqi et. al.~\cite{9139854} studied the performance and pitfalls of several CUDA synchronization methods for reduction operations.
Sinclair et. al.~\cite{8167781} presented a benchmark suite to measure the performance of synchronization primitives for different coherence protocols and consistency models.

Stream-K~\cite{streamk} is a GeMM implementation that improves the utilization of SMs of a GPU by dividing the workload among all SMs.
However, Stream-K is not straightforward to apply to computations other than GeMMs. 
In contrast, \libname{} fits thread blocks of multiple kernels in each wave and is applicable to any tile based computations.

\section{Conclusion}
State-of-the-art ML models consist of thousands of individual computations that are executed on one or more GPUs.
However, these models under-utilize the GPUs because individually each of these computations cannot completely utilize a GPU and these models largely consists of dependent computations.
In this paper, we presented \libname{}, a framework for fine-grained synchronization of tiles of dependent computations.
By synchronizing only, the dependent tiles, our framework allows concurrent execution of independent tiles, thus improving the utilization of GPU.
Our experiments show that synchronizing computations of existing machine learning models using \libname{} can reduce inference times of these models significantly.   

\appendix
\section{Artifact Appendix}
The artifact~\cite{Jangda2023} contains \libname{} CUDA implementation and scripts to reproduce all of our results.
The artifact provides both a Dockerfile, which contains all prerequisites installed, and source code.
Latest source code is available at \url{https://github.com/microsoft/cusync}.
The artifact reproduces Figure~\ref{fig:results:gpt-3},~\ref{fig:results:resnet}, and ~\ref{fig:results:e2e} in Section~\ref{sec:eval}.

\spara{System} We executed our experiments on a NVIDIA DGX-2 system containing 8 NVIDIA Tesla V100 GPUs connected using NVLINK.
Our experiments will run on any system with a GPU, however, the end-to-end inference results in Figure~\ref{fig:results:e2e} might not be reproducible on another system.

\spara{Extract Artifact} Download the artifact from ~\cite{Jangda2023} and extract the zip file.

\begin{lstlisting}[basicstyle=\small\ttfamily]
unzip cusync-cgo-24.zip
cd cusync-cgo-24
\end{lstlisting}

\subsection{Docker Container}
To run artifact inside a Docker container follow these steps:

\spara{Install docker} Install docker engine by following steps on \url{https://docs.docker.com/engine/install/ubuntu/}.

\spara{Install NVIDIA Container Toolkit} Install NVIDIA Container Toolkit by following steps on \url{https://docs.nvidia.com/datacenter/cloud-native/container-toolkit/latest/install-guide.html}.

\spara{Create Container} Create docker container using the Dockerfile, start the container, and cd to the directory:

\begin{lstlisting}[basicstyle=\small\ttfamily]
docker build -t cusync-cgo-24 .
docker run -it --gpus all cusync-cgo-24
cd /cusync
\end{lstlisting}

\spara{Check PyTorch and CUDA Install}: Check if torch supports CUDA if \texttt{torch.cuda.is\_available()} returns \texttt{True}:

\begin{lstlisting}[basicstyle=\small\ttfamily]
python
>>> import torch
>>> torch.cuda.is_available()
True
\end{lstlisting}

\subsection{Running Natively}
We can also run code natively, which requires installing all dependencies.
These steps can be ignored if using docker in above steps.

\spara{Linux Installation} We recommend using Ubuntu 22.04 as the Linux OS. We have not tested our artifact with any other OS but we believe Ubuntu 20.04 and 23.04 should also work.

\spara{Install Dependencies} Execute following commands to install dependencies.

\begin{lstlisting}[basicstyle=\small\ttfamily]
sudo apt update 
sudo apt install gcc linux-headers-$(uname -r)\ 
make g++ git python3 wget\
unzip python3-pip build-essential cmake
\end{lstlisting}

\spara{Install CUDA} We need to install CUDA before proceeding further. In our experiments we used CUDA 12.2 on Ubuntu 
\\
\\
22.04. CUDA 12.2 toolkit can be downloaded from \url{https://developer.nvidia.com/cuda-12-1-0-download-archive}.
After installing CUDA, set \texttt{nvcc} and CUDA paths.

\begin{lstlisting}[basicstyle=\small\ttfamily]
export PATH="/usr/local/cuda/bin:$PATH"
export LD_LIBRARY_PATH=
"/usr/local/cuda/lib64:$LD_LIBRARY_PATH"
\end{lstlisting}

\spara{Check CUDA Installation} To check CUDA installation, run \texttt{nvidia-smi} and it should print all GPUs in the system. 
Otherwise there is a problem with the CUDA installation.
\spara{Install Pytorch}: Install PyTorch using pip.

\begin{lstlisting}[basicstyle=\small\ttfamily]
sudo pip3 install torch torchvision torchaudio
\end{lstlisting}

\spara{Check Pytorch CUDA Install}: Check if torch supports CUDA if \texttt{torch.cuda.is\_available()} returns \texttt{True}:

\begin{lstlisting}[basicstyle=\small\ttfamily]
python
>>> import torch
>>> torch.cuda.is_available()
True
\end{lstlisting}

\spara{Obtain source code}
The source code can be downloaded from \cite{Jangda2023}. 
Latest source code is available from cuSync repository and CGO AE branch:
\begin{lstlisting}[basicstyle=\small\ttfamily]
git clone --recurse-submodules \
https://github.com/microsoft/cusync
cd cusync
git checkout cgo-24-ae
\end{lstlisting}

\subsection{Functionality and Reusability}
The \texttt{README.md} contains instructions of how code can be compiled to other NVIDIA GPU architectures, an example and test cases.
The functionality can be checked by executing these test cases.
To run tests execute:
\begin{lstlisting}[basicstyle=\small\ttfamily]
make tests -j
\end{lstlisting}
If all tests passes then we are ready for reproducing results.

\subsection{Reproduce Results}
We will now reproduce our main results of Figure~\ref{fig:results:gpt-3},~\ref{fig:results:conv}, and ~\ref{fig:results:e2e}.
All commands should be executed in the \texttt{cusync} directory.

\spara{Large Language Model Inference Results}
[Time 60 mins]
Following commands will run all experiments to gather the results

\begin{lstlisting}[basicstyle=\small\ttfamily]
cd src/ml-bench/volta_transformer
python3 eval_llm.py mlp gpt3
python3 eval_llm.py attention gpt3
python3 eval_llm.py mlp llama
python3 eval_llm.py attention llama
python3 allreduce_times.py
\end{lstlisting}

\spara{Computer Vision Inference Results}
[Time 60 mins]
Following commands will run all experiments to gather results for Figure~\ref{fig:results:conv}.
\begin{lstlisting}[basicstyle=\small\ttfamily]
cd src/ml-bench/volta_conv2d
python3 eval_conv.py resnet
python3 eval_conv.py vgg
\end{lstlisting}

\spara{Generate Plots}
[Time 5 mins]
Generate all Figures by running below commands:
\begin{lstlisting}[basicstyle=\small\ttfamily]
cd src/ml-bench/plots
make -j
\end{lstlisting}
The current directory will have figures as PDF and they can be checked against figures in the paper.

\bibliographystyle{IEEEtran}
\bibliography{paper}
\end{document}